\documentclass{llncs}

\usepackage[utf8]{inputenc}

\usepackage[T1]{fontenc}
\usepackage{flushend}
\usepackage{amsmath}
\usepackage{bm}

\PassOptionsToPackage{hyphens}{url}
\usepackage{hyperref}

\usepackage[hyphenbreaks]{breakurl}

\usepackage[detect-weight=true, detect-family=true,per-mode=symbol-or-fraction,binary-units=true]{siunitx}
\DeclareSIUnit\sample{Sa}
\DeclareSIUnit\cycle{cycle}
\DeclareSIUnit\instruction{instruction}
\DeclareSIUnit\core{core}

\usepackage{subfig}

\usepackage{tabularx}

\usepackage{booktabs}

\usepackage{graphicx}

\usepackage{listings}

\title{Introducing the \emph{Arm-membench} Throughput Benchmark}

\author{Cyrill Burth\orcidID{0009-0001-6030-3201} \and Markus Velten\orcidID{0000-0002-2730-0308} \and Robert Sch\"{o}ne\orcidID{0009-0003-0666-4166}\institute{ZIH, CIDS, TU Dresden, Dresden, 01062, Germany\\ \email{cyrill.burth@mailbox.tu-dresden.de}\\ \email{\{markus.velten,robert.schoene\}@tu-dresden.de}}}

\newcommand{\figref}[1]{\textcolor{red}{Figure~\ref{fig:#1}}}
\newcommand{\tabref}[1]{\textcolor{red}{Table~\ref{tab:#1}}}
\newcommand{\secref}[1]{\textcolor{red}{Section~\ref{sec:#1}}}
\newcommand{\secrefrange}[2]{Sections~\ref{sec:#1} to \ref{sec:#2}}
\newcommand{\lstref}[1]{\textcolor{red}{Listing~\ref{lst:#1}}}

\begin{document}

	\maketitle

\begin{abstract}
Application performance of modern day processors is often limited by the memory subsystem rather than actual compute capabilities.
Therefore, data throughput specifications play a key role in modeling application performance and determining possible bottlenecks.
However, while peak instruction throughputs and bandwidths for local caches are often documented, the achievable throughput can also depend on the relation between memory access and compute instructions.
In this paper, we present an Arm version of the well established x86-membench throughput benchmark, which we have adapted to support all current SIMD extensions of the Armv8 instruction set architecture.
We describe aspects of the Armv8 ISA that need to be considered in the portable design of this benchmark.
We use the benchmark to analyze the memory subsystem at a fine spatial granularity and to unveil microarchitectural details of three processors: Fujitsu A64FX, Ampere Altra and Cavium ThunderX2.
Based on the resulting performance information, we show that instruction fetch and decoder widths become a potential bottleneck for cache-bandwidth-sensitive workloads due to the load-store concept of the Arm ISA.
\end{abstract}

\section{Introduction}
\label{sec:intro}
Multicore designs and increasing widths of \textbf{S}ingle \textbf{I}nstruction \textbf{M}ultiple \textbf{D}ata (SIMD) extensions have increased the peak performance significantly.
\renewcommand{\thefootnote}{}
\footnotetext{\scriptsize This preprint has not undergone peer review or any post-submission improvements or corrections. The Version of Record of this contribution is published in Parallel Processing and Applied Mathematics 2024, and is available online at \url{https://doi.org/10.1007/978-3-031-85697-6_7}}
\renewcommand{\thefootnote}{\arabic{footnote}}
However, memory performance did not scale with the compute performance~\cite[Figure 1.23]{Henessy_Patterson_Computerarchitecture_6th}, leading to the development of mitigation techniques such as caches, which establish a memory hierarchy with varying access latencies and bandwidths, and prefetchers, which fetch data to these caches in advance.
Hence, cache sizes, latency, and bandwidth become crucial factors in modeling application performance~\cite{ecm,multibsp}.
While the L1 data (L1d) cache is usually designed to match the maximum achievable throughput of the SIMD execution units, higher memory levels will not be able to match this performance~\cite{MolkaNehalem,MolkaHaswell,VeltenRomeCLX}.
Developments such as the transition towards \textbf{N}on \textbf{U}niform \textbf{M}emory \textbf{A}ccess (NUMA) designs make memory access patterns even more complex and demand careful data partitioning and allocation.
This creates a demand for tools capable of revealing differences in various memory access patterns across the different memory locations and support a way to measure their performance.

This paper presents a port of the established x86-membench throughput benchmark to the Armv8 \textbf{I}nstruction \textbf{S}et \textbf{A}rchitecture (ISA)~\cite{arm_isa_manual} with support for both SIMD extensions: NEON and the \textbf{S}calable \textbf{V}ector \textbf{E}xtension (SVE).
We present performance measurements for three different Arm server processors, proving that our benchmark functions correctly and highlighting important microarchitectural differences.

\section{Related Work}
\label{sec:rel}
Since memory performance is a crucial limitation in computing systems, various benchmarks have been developed to measure effective bandwidth properties.
Of these, \textit{STREAM}~\cite{stream} is the most prevalent.
While it is easy to use, parallel, and not only reflects the memory bandwidth but also floating point performance, it has major disadvantages:
First, the measured performance significantly depends on the used compiler and compiler flags.
Second, the OpenMP overhead is most often too high to get results for datasets that fit into the lower cache levels~\cite[Section 3]{schoene_stream_cache_2009}.
Alternative benchmarks attempt to address some of these issues.
One of them is \textit{likwid-bench} by Treibig et al. \cite{likwid_bench}.
It uses a domain specific language, avoiding the performance influence of the compiler, to define workloads that can be used to measure the throughput of x86 and Arm systems.
It uses a more precise timer and provides more accurate results than STREAM, making it more suitable for the analysis of cache properties.
However, users cannot freely select the data used, but only the data type.
Both data type and processed data influence power consumption~\cite{Molka_energy,Schoene_2019_SKL}, and subsequently, in power-constrained scenarios, performance.
Furthermore, the entire memory hierarchy cannot be analyzed with a single run.
The \textit{x86-membench} suite by Molka~\cite{MolkaDiss} can be used for a detailed analysis of microarchitectures.
However, as the name indicates, it targets only x86-based processors.
In this work, we extend a benchmark of this suite to run on the Arm architecture.

\section{Background}
\label{sec:bg}
\subsection{The Arm Instruction Set}
\label{sec:bg-arm}
In this work, we present a benchmark that is specifically designed for the Armv8 ISA.
This requires the consideration of some of Armv8's unique features.

With a load-store architecture~\cite[Chapter 1]{Henessy_Patterson_Computerarchitecture_6th}, operands cannot be directly accessed from memory but need to be explicitly loaded to a register first.
This increases the number of instructions and subsequently pressure on the front end and \textbf{o}ut-\textbf{o}f-\textbf{o}rder (OoO) resources.
To address this issue, data could be held in registers for longer to increase the computation per memory transfer ratio, but this depends on the workload and is not always feasible.

The Armv8 ISA specifies a mandatory vector extension with a register width of \SI{128}{\bit}, called NEON.
NEON instructions can operate on a variety of differently sized data from \SIrange{8}{64}{\bit} per element and support all common arithmetic and logic operations.
The optional Scalable Vector Extension uses variable vector lengths from \SIrange{128}{2048}{\bit}.
This allows implementations to use different register sizes without the need to adjust or recompile the source code.

\subsection{x86-membench}
\label{sec:bg-x86memb}
\textit{x86-membench}~\cite{MolkaDiss} is a suite of benchmarks that enables users to determine the throughput of compute instructions, bandwidth measured with data access instructions, and latency properties for the entire memory hierarchy.
The included benchmarks use assembly-routines to counteract compiler effects.
A configuration file for each benchmark offers fine-grained controls.
For instance, specific instructions, in particular those introduced with SIMD extensions, can be selected when measuring execution throughput.
Cache coherence states can be selected when analyzing memory access latencies.
Depending on the benchmark, the usage of one or more cores can be specified.
Within a single measurement run, the entire memory hierarchy can be analyzed, from the L1d cache to main memory.
The benchmark includes support for hugepages and memory pinning, unlike, for example, STREAM, which relies on external tools like \texttt{numactl} for the latter purpose.

The x86-membench throughput benchmark~\cite[Section 3.5.4]{MolkaDiss} measures the time it takes to process a loop with a fixed number of arithmetic or logic instructions accessing predefined data.
Time is precisely measured using the \texttt{rdtsc} instruction, which reads the (fixed) time stamp counter, with serialization using \texttt{lfence} and \texttt{mfence} instructions.
If all data resides in registers, the peak throughput for this instruction can be determined.
If data has to be loaded from the higher levels of the memory hierarchy, throughput can be limited by the bandwidth of the respective memory level.
Thus, the results reflect the maximum throughput a given instruction can achieve when reading operands from memory, as long as no additional bottlenecks, e.g., in the front end, exist.
The overhead of the loop itself is statically analyzed and can be considered in the calculation of the results.

Avoiding the occurrence of any denormal numbers is ensured by initializing the buffer through a series of a user-defined number, the reciprocal of that number, as well as the additive inverse of these two.

\section{Benchmark Design}
\label{sec:design}
We retain the general structure of the x86-membench throughput benchmark for the Armv8 port.

We read the generic timer register \texttt{CNTVCT} to get time stamps with low overhead.
It is incremented at a constant rate, exposed to the user through the \texttt{CNTFRQ} register.
Serialization is ensured by employing two types of barriers before generating the time stamp.
First, a Data Synchronization Barrier (\texttt{DSB SY}) garantuees the execution and visibility of preceding load/store operations.
This is followed by an Instruction Synchronization Barrier (\texttt{ISB}), enforcing that all subsequent instructions are fetched again.

Unlike the width of NEON registers, that of SVE is not known at compile time.
Hence, the benchmark discovers it at runtime via the \texttt{INC\{D/W/H/B\}} instruction, enabling a vector-length-agnostic SVE implementation.

As discussed in \secref{bg-arm}, the benchmark requires additional load instructions, which put more pressure on the front end and OoO resources.
If the front end cannot fetch and decode sufficient instructions per cycle, execution units may idle, limiting throughput by fetch and decode performance rather than load/store capabilities.
Therefore, we also measured the raw load throughput using a loop with only \texttt{LD1} or \texttt{LD2D} instructions and performing no arithmetic operations.
These values represent the peak throughput that can be achieved by the given architecture.
We substituted arithmetic instructions, e.g., \texttt{FADD}, with \texttt{NOP}s.
These instructions need to be fetched, decoded and committed by the OoO engine but do not allocate resources in execution units.
As a result, execution time primarily depends on the time required for the load instructions and the width of the fetch, decode, and commit units, as they need to provide a sufficient amount of instructions per cycle to keep the load/store units busy.
Ideally, these throughput results are equal to those of the arithmetic instructions.
If bottlenecks exist from the execution units' side, this will be reflected by a lower throughput of arithmetic instructions compared to \texttt{NOP}s.

We therefore designed the benchmark carefully with a low number of instructions.
For example, we use the instructions \texttt{LD1} (NEON) or \texttt{LD2D} (SVE) to load data to multiple registers simultaneously.
The number of instructions could be minimized further by using a post-increment to increment the address pointer with the NEON \texttt{LD1} instruction.
The SVE \texttt{LD2D} instruction lacks the post-increment feature.
Instead, the benchmark offers two implementations for each SVE kernel: one with offset addressing and the other with manual memory pointer increment.
Possible implementations for NEON are shown in \lstref{nopostincr} and \lstref{postincr}, SVE implementations using the \texttt{LD2D} instruction are not depicted but follow the same structure.
\begin{figure}[b]
\centering
\begin{minipage}[t]{.4\textwidth}
\begin{lstlisting}[caption=Workload with a manual pointer increment,label=lst:nopostincr,numbers=left,numberstyle=\tiny,frame=tlrb]
LD1 {v16.2d-v19.2d},[X0]
ADD X0, X0, #256
LD1 {v20.2d-v23.2d},[X2]
ADD X2, X2, #256
FADD v0.2d, v0.2d, v16.2d
<+6x FADD COMMANDS>
FADD v7.2d, v7.2d, v23.2d
\end{lstlisting}
\end{minipage}
\hspace{1cm}
\begin{minipage}[t]{.4\textwidth}
\begin{lstlisting}[caption=Workload with post-increment,label=lst:postincr,numbers=left,numberstyle=\tiny,frame=tlrb]
LD1 {v16.2d-v19.2d},[X0],#64
LD1 {v20.2d-v23.2d},[X0],#64
FADD v0.2d, v0.2d, v16.2d
<+6x FADD COMMANDS>
FADD v7.2d, v7.2d, v23.2d
\end{lstlisting}
\end{minipage}
\end{figure}

Both approaches compute the same floating point addition (\texttt{FADD}, lines 3, 5 and 7) and operate on the same memory addresses.
In \lstref{nopostincr}, the pointer is incremented manually (lines 2 and 4), whereas the \texttt{LD1} instruction in \lstref{postincr} is annotated with the correct increment (lines 1 and 2).
To minimize dependencies between memory accesses in the manual increment implementation, the address pointer is duplicated with an offset across four general-purpose registers (\texttt{X0} and \texttt{X2} are shown in \lstref{nopostincr}), thereby increasing the increment from \SI{64}{\byte} to \SI{256}{\byte}.
It is not possible to use multiple address pointers with the post-increment, as this only allows \SI{32}{\byte} or \SI{64}{\byte} increments~\cite[Section LD1 (multiple structures)]{arm_isa_manual}.

\begin{figure}[bth]
\centering
\includegraphics[height=0.14\textheight]{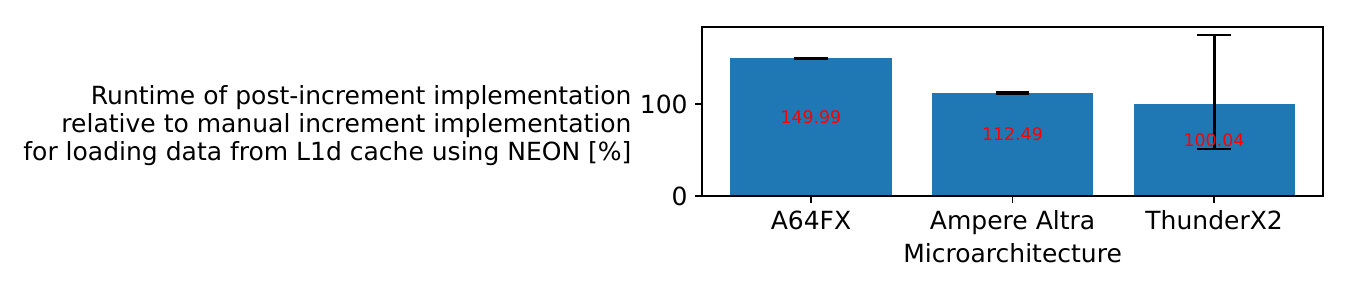}
\caption{Relative L1d cache performance of post-increment implementation.}
\label{fig:post-incr-influence}
\end{figure}

\figref{post-incr-influence} depicts the runtime overhead of post-increment over our manual increment implementation for loading the same amount of data from L1d cache using NEON vector registers.
The presented results show the arithmetic mean over one hundred measurements.

From the A64FX manual \cite{a64fx_manual} it can be concluded that for NEON instructions with post-increment an extra $\mu$OP is created which executes on the EAG\{A/B\} pipelines like the load execution flow itself.
Upon examining performance counters on the Ampere Altra, particularly \texttt{dp\_spec}, which tracks integer data processing operations, we can infer that the post-increment operations are not executed by the integer data processing pipelines, unlike the manual increment.
The post-increments are most likely executed on the address generation units connected to the load/store pipeline, similar to the A64FX.
To achieve peak performance the load/store units must remain fully utilized at all times and provide, e.g., two instructions per cycle for the architectures discussed in this paper.
It seems reasonable that imposing additional instructions onto already fully utilized execution units would degrade performance.
On the ThunderX2, performance counters revealed that most of the time, the processor stalls due to the front end. 
However, using post-increment did not influence performance, though measurements exhibited large outliers in both cases.
A limited number of available performance counters prohibited further analysis of this behavior.

Thus, we use the manual increment approach with multiple address pointers for NEON, also avoiding potential restrictions imposed by the maximum size for the post-increment.
A comparison of SVE using a manual increment and an offset addressing had to be omitted due to space constraints.
The SVE implementation uses offset addressing since it proved to be the most beneficial, except when using only \texttt{LD2D} load instructions.
In those cases, we observed a reproducible drop in the L1d cache bandwidth for the offset addressing mode and decided to use manual increment, as it provided equal performance with less noise in the lower cache levels.
However, the presented benchmark offers different kernels for SVE, NEON, and general purpose registers, including versions with and without manual increment, allowing the choice of the best addressing modes for other architectures.

\section{Methodology \& Test Systems}
\label{sec:method}
At each memory level, we measure the performance of the NEON and/or SVE extension(s), depending on the processor.
We use \texttt{FADD} with double-precision elements as arithmetic instruction within the measurement routine.
Alternatively, we utilize the \texttt{NOP} instruction to prevent the SIMD units from becoming a bottleneck.
In future works other instructions can be configured.

In the presented results, we depict the cumulative mean over one hundred internal repetitions of the benchmark.
Additionally, we used the arithmetic mean of four consecutive memory accesses for plots that show aggregated values.
We report standard deviations for these plots.

The bandwidths of, in particular, caches, are often publicly documented by the manufacturer.
The main memory performance was determined through the number of channels and the frequency of the available memory configuration.
However, achievable performance often differs from the theoretical peak.
Whenever available, our own measurements will be compared to those from other scientific publications to ensure the correctness of our benchmark.

We evaluate the benchmark on three Arm processors with different microarchitectures.
In all presented measurements transparent hugepages are used.
\tabref{hw_specs} lists key specifications of the three systems.
In \secrefrange{a64fx}{thunderx2}, we describe additional details that are relevant to this work.

\begin{table}[!h]
\centering
\caption{Hardware and software specification for the test systems in use.}
\label{tab:hw_specs}
{\setlength{\tabcolsep}{0.2em}
\footnotesize{
\begin{tabular}{rrrr}
 \toprule
 &\textbf{Fujitsu}&\textbf{Ampere}&\textbf{Marvell}\\
 &\textbf{A64FX}&\textbf{Altra Q80-30}&\textbf{ThunderX2 CN9975}\\
 \midrule
 Sockets \texttimes\,Cores/Socket & 1 \texttimes\,48 & 1 \texttimes\,80 & 2 \texttimes\,28 \\
 Frequency & \SI{1.8}{\giga\hertz} & \SI{3}{\giga\hertz} & \SI{2}{\giga\hertz} \\
 SMT & -- & -- & $4\times$ \\
 ISA & Armv8.2-A & Armv8.2-A & Armv8.1\\
 SIMD ext. (width) & SVE (\SI{512}{\bit}) & NEON (\SI{128}{\bit}) & NEON (\SI{128}{\bit}) \\
 Decoder width & 4 instr./cycle & 4 instr./cycle & 4 instr./cycle \\
 \midrule
 L1d Cache per Core & \SI{64}{\kibi\byte} & \SI{64}{\kibi\byte} & \SI{32}{\kibi\byte} \\
 L1d Cache B/W per Core& \SI{230.4}{\giga\byte\per\second} & \SI{96}{\giga\byte\per\second} & \SI{64}{\giga\byte\per\second} \\
 L2 Cache & \SI{8}{\mebi\byte} per CMG & \SI{1}{\mebi \byte} per core & \SI{256}{\kibi\byte} per core \\
 L3 Cache & -- & \SI{32}{\mebi\byte} & \SI{28}{\mebi\byte} \\
 DRAM Type, Channel & HBM2, 4 & DDR4-3200, 8 & DDR4-2666, 8 \\
 DRAM Amount & \SI{32}{\gibi\byte} & \SI{512}{\gibi\byte} & \SI{128}{\gibi\byte} \\
 DRAM B/W per Socket & \SI{921.6}{\giga\byte\per\second} & \SI{204.8}{\giga\byte\per\second} & \SI{170.5}{\giga\byte\per\second} \\
 \midrule
 Linux OS, Kernel & Rocky 8.9, 4.18 & Rocky 8.6, 4.18 & Ubuntu 22.04.1, 5.15 \\
 \midrule
 Documentation & \cite{a64fx_manual} & \cite{ampere_altra_datasheet}, \cite{arm_neoverse_n1_manual}, \cite{neoverse_n1_white} & N.A.\\
 \bottomrule
\end{tabular}
}}
\end{table}

\subsection{Fujitsu A64FX}
\label{sec:a64fx}
The Fujitsu A64FX is used in the FUGAKU HPC system~\cite{A64FX-ECM,A64FX_HPC_APPs}.
It was the first processor to implement SVE, using \SI{512}{\bit} wide registers \cite{a64fx_manual}.
The processor itself is structured in four \textbf{C}ore \textbf{M}emory \textbf{G}roups (CMG), each forming its own NUMA node and holding up to 12+1 cores, for a total of 52 cores per CPU.
48 cores are dedicated to user workloads, whereas the four optional assistant cores are not exposed through the operating system.

Without assumptions regarding the instruction mix, the instruction buffer can send up to four instructions to the decode units each cycle \cite{a64fx_manual}.
Each core has two load/store units, each capable of loading \SI{512}{\bit} per operation, for a total of \SI{1024}{\bit} or \SI{128}{\byte}, served by the L1d cache per cycle.
The L2 cache is implemented as two cache banks \cite{okazaki2020supercomputer} and can serve loads to the L1d at a maximum of \SI{64}{\byte\per\cycle}, limited to \SI{512}{\byte\per\cycle} per CMG for reads \cite{a64fx_manual}.
With a clock frequency of \SI{1.8}{\giga\hertz}\footnote{As described in \url{https://www.nhr.kit.edu/userdocs/ftp/hardware/}}, this leads to a theoretical peak bandwidth for loads of \SI{230.4}{\giga\byte\per\second} from the L1d cache and \SI{115.2}{\giga\byte\per\second} from the L2 cache.

As main memory \SI{32}{\gibi\byte} of on-package \textbf{H}igh \textbf{B}andwidth \textbf{M}emory 2 (HBM2) is used.
Each CMG is connect to one \SI{8}{\gibi\byte} HBM2 stack.
The bandwidth of one HBM2 stack to the local CMG's L2 cache is documented with \SI{128}{\byte\per\cycle}, leading to a per-stack bandwidth of \SI{230.4}{\gibi\byte\per\second} and a combined \SI{921.6}{\gibi\byte\per\second} for all CMGs.

\subsection{Ampere Altra}
\label{sec:altra}
The Ampere Altra uses cores based on the Arm Neoverse-N1~\cite{arm_neoverse_n1_manual} design for server CPUs.
Our test system was an Ampere Altra Q80-30 with 80 cores that run at a frequency of \SI{3}{\giga\hertz}.
The \SI{32}{\mebi\byte} L3-cache is shared among all cores.
Each core offers two \SI{128}{\bit} read paths from the L1d cache, providing a bandwidth of \SI{96}{\giga\byte\per\second}.
A 4-way decoder in each core enables a sustained maximum performance of 4 instructions per cycle \cite{neoverse_n1_white}.

\subsection{Marvell ThunderX2}
\label{sec:thunderx2}
The ThunderX2\footnote{Unfortunately, we were unable to find any publicly available first-party documentation for this processor or the microarchitecture.} was the first Arm-based processor to enter the Top500 list in 2018~\cite{thunderx2_astra_hpc}\footnote{\url{https://top500.org/system/179565/}}.
The L3 cache is implemented through \SI{1}{\mebi\byte} slices per core, resulting in \SI{28}{\mebi\byte} of shared system level cache.
Two \SI{128}{\bit} load paths towards the L1d cache are available per core, achieving a theoretical bandwidth of \SI{64}{\giga\byte\per\second} at a frequency of \SI{2}{\giga\hertz}.
Each core is able to decode up to four instructions per cycle and send them to the scheduler and dispatch unit.

\begin{figure}[b!]
  \centering
  \includegraphics[height=0.17\textheight]{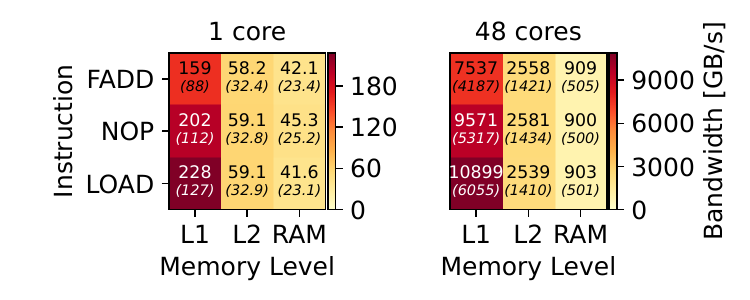}
  \caption{A64FX: Throughput of different SVE instructions for all memory levels using a single core and all 48 cores in [GB/s] and \textit{([B/cycle])}. The standard deviation is $<$ \SI{1}{\percent} for all cases.}
  \label{fig:a64fx_single_multi}
\end{figure}
\section{Benchmark Evaluation}
\label{sec:experiment}
\subsection{Fujitsu A64FX}
\label{sec:results_a64fx}
The SVE measurements for a single core and all 48 cores, showing the complete memory hierarchy of the A64FX, are depicted in \figref{a64fx_single_multi}.
We measured a L1d cache throughput of \SI{69}{\percent} of the theoretical peak for the \texttt{FADD} case.
If we substitute \texttt{FADD} instructions with \texttt{NOP}s, \SI{88}{\percent} can be achieved.
When using only the \texttt{LD2D} load instruction, we measured \SI{99}{\percent}.
The L1d cache bandwidth can only be fully saturated when using only load instructions.
We assume that the A64FX's front end and potentially OoO resources may not be able to process enough instructions each cycle when not just using load instructions to keep the load/store units busy, as described in \secref{design}.
The L2 cache throughput is slightly impacted by the choice of instruction mix, but at a much smaller level with differences of \SI{0.9}{\giga\byte\per\second}.
In all cases approx. \SIrange{50}{51}{\percent} of the theoretical peak was measured.
Presumably, the data distribution over both L2 cache banks is not optimal; the analysis is left for future work.

\begin{figure}[b]
\centering
\includegraphics[height=0.17\textheight]{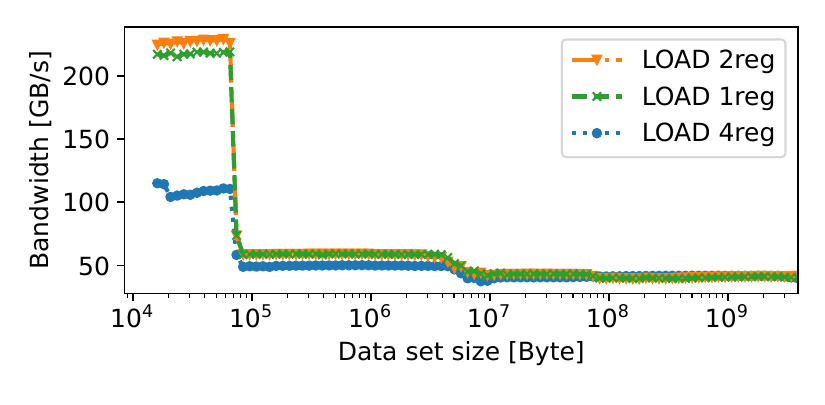}
\caption{A64FX: Bandwidth of SVE with different numbers of registers loaded per instruction using \texttt{LD1D}, \texttt{LD2D} and \texttt{LD4D} for loading one, two or four registers.}
\label{fig:a64fx_sve-regs}
\end{figure}

It is possible to load multiple SVE registers with the same instruction, e.g. with \texttt{LD2D} or \texttt{LD4D}, thus reducing the overall instruction footprint.
However, as is shown in \figref{a64fx_sve-regs}, peak performance can only be achieved when loading two registers with a single instruction.
Loading four registers at once leads to a worse performance compared to loading only a single register per instruction.
The reason lies within the implementation of the A64FX \cite{a64fx_manual}.
The \texttt{LD\{2/3/4\}D} instruction loads interleaved data elements into two, three, or four consecutive registers. 
Loading the registers is accomplished through separate execution flows, where each flow comprises one or more memory access flows.
One memory access flow fetches \SI{128}{\byte} from the L1d cache and loads the elements into one register.
If the data is stored consecutively in memory, a single memory access flow for each execution flow loads a full register in the case of \texttt{LD1D} and \texttt{LD2D}. 
However, in the case of \texttt{LD3D} or \texttt{LD4D}, not all elements of a single register are contained within the \SI{128}{\byte} fetch window, necessitating an additional memory access flow.
This can also be observed with \texttt{perf}.
A workload that has elements consecutively stored in L1d cache and loads them with \texttt{LD\{3/4\}D} instructions has twice as many L1d cache accesses compared to using \texttt{LD\{1/2\}D} instructions.
The scenario assumes that the data to be loaded is \SI{128}{\byte} aligned.
Our benchmark supports custom memory alignment and provides kernels for all implementations. 
 
For the \texttt{FADD} loop with all 48 cores, we measured \SI{68}{\percent} of the theoretical peak (\figref{a64fx_single_multi}).
The measurements with only load or \texttt{NOP} instructions achieve similar performance compared to the single core case with \SI{99}{\percent} and \SI{87}{\percent}, respectively.
All cases reach $\approx48\times$ the single core performance.
The same impact of the instruction mix on the L2 cache bandwidth can be observed.
In all cases the L2 cache did not scale perfectly linearly, up to $\approx44\times$ the single core performance.

\begin{figure}[t]
\centering
\includegraphics[height=0.17\textheight]{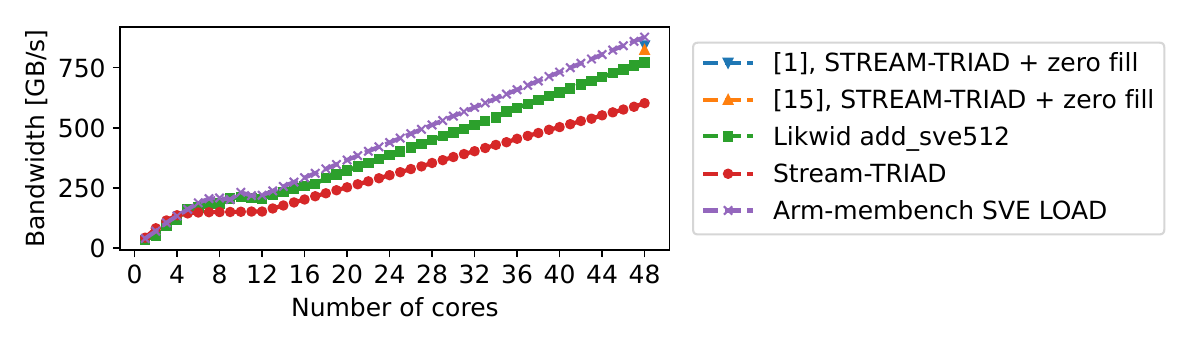}
\caption{A64FX: HBM2 scaling behavior of STREAM TRIAD and the Arm-membench throughput benchmark on one socket, starting with cores in CMG 0. Values from \protect\cite{A64FX-ECM} and \protect\cite{A64FX_HPC_APPs} for 48 cores shown as reference. Both use STREAM TRIAD with zero fills.}
\label{fig:a64fx_main_memory}
\end{figure}

In order to evaluate the main memory results of our benchmark, we compare it with the STREAM benchmark, as depicted in \figref{a64fx_main_memory}.
Our benchmark achieves a bandwidth of \SI{909}{\giga\byte\per\second} on all cores, which is \SI{99}{\percent} of the theoretical peak.
Poenaru et al. \cite{A64FX_HPC_APPs} presented STREAM TRIAD results of up to \SI{824}{\giga\byte\per\second}, while Alappat et al. measured \SI{841}{\giga\byte\per\second} \cite{A64FX-ECM}, using the Fujitsu C Compiler (FCC), which automatically utilizes the zero fill operations.
The zero fill instruction \texttt{DC ZVA} zeros a specific cacheline and puts it directly into the L2 cache when a cache write miss occurs, enabling the processor to load it directly from the L2.
If no zero fill instruction is used, e.g., by using the GNU project C Compiler (GCC), $\approx$\SI{600}{\giga\byte\per\second} can be achieved, see \figref{a64fx_main_memory} and \cite{A64FX-ECM,A64FX_HPC_APPs}.
The FCC compiler was not available on our test system at the time of writing.
The measurements from our benchmark surpass the STREAM measurements with the FCC compiler shown in \cite{A64FX-ECM,A64FX_HPC_APPs}.
This is expected since our measurement routine is written in assembly and does not perform any writes; therefore, the utilization of the zero fill operation has no influence.
For both benchmarks, a similar scaling of HBM2 bandwidth is observed.
With six cores of a single CMG we achieve bandwidth saturation, with \SI{227}{\giga\byte\per\second} for our benchmark, which is identical to the measurements presented by Alappat et al. \cite{A64FX-ECM}, and $\approx$\SI{151}{\giga\byte\per\second} for STREAM TRIAD.
Utilizing multiple CMGs leads to a near linear increase in performance.
This is expected and was already shown by Alappat et al. \cite{A64FX-ECM}, since the bandwidth is calculated by the amount of data read over the time it took the slowest thread to complete.

\begin{figure}[b!]
  \centering
  \includegraphics[height=0.17\textheight]{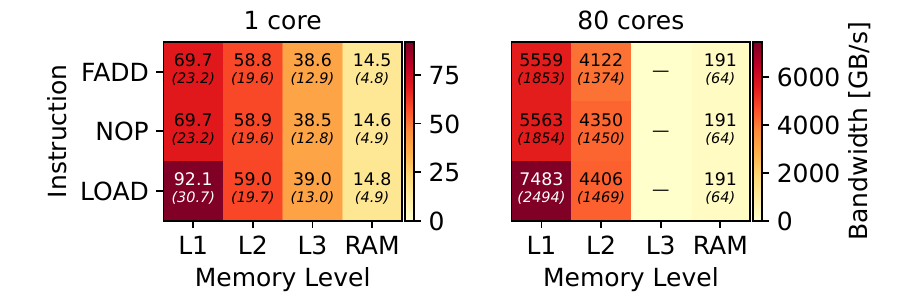}
  \caption{Ampere Altra: Throughput of different NEON instructions for all memory levels using a single core and all 80 cores in [GB/s] and \textit{([B/cycle])}. Multicore L3 accesses could not be distinguished due to small L3 size (denoted as '--'). The standard deviation is $<$ \SI{1}{\percent} for all cases except multicore L1 LOAD ($\approx$ \SI{3}{\percent}).}
  \label{fig:altra_single_multi}
\end{figure}

\subsection{Ampere Altra Q80-30}
\label{sec:results_aa}

\figref{altra_single_multi} shows measurements using the NEON extension on the Ampere Altra Q80-30.
We measured a L1d throughput of \SI{73}{\percent} of the theoretical peak for a single core with \texttt{FADD} instructions.
Using only load instructions and \texttt{NOP} instead of \texttt{FADD}, the processor achieves \SI{96}{\percent} and \SI{73}{\percent} of the peak performance.
This may suggest a congested front end and potentially limiting OoO capabilities, similar to the observations made on the A64FX (\secref{results_a64fx}).
The L2 and L3 cache throughput are not impacted by the choice of the instruction mix with a difference of \SI{0.5}{\giga\byte\per\second} between the lowest and highest throughput, reaching around \SI{59}{\giga\byte\per\second} and \SI{39}{\giga\byte\per\second}, respectively.

The L1d cache throughput scales linearly with the number of cores, reaching $\approx80\times$ the throughput of a single core.
We measured the lowest throughput for the L2 cache using the \texttt{FADD} loop, with $\approx70\times$ the throughput of a single core.
The highest throughput was achieved using only load instructions, reaching $75\times$ the throughput of a single core and $\approx74\times$ with \texttt{NOP}.
The L3 throughput cannot be properly distinguished, as the L3 is relatively small (\SI{32}{\mebi\byte}) compared to the all-core L1d + L2 capacity (\SI{85}{\mebi\byte}).
Main memory performance was measured with \SI{93}{\percent} of its peak performance.

\subsection{Marvell ThunderX2}
\label{sec:results_thx2}

\begin{figure}[b!]
  \centering
  \includegraphics[height=0.17\textheight]{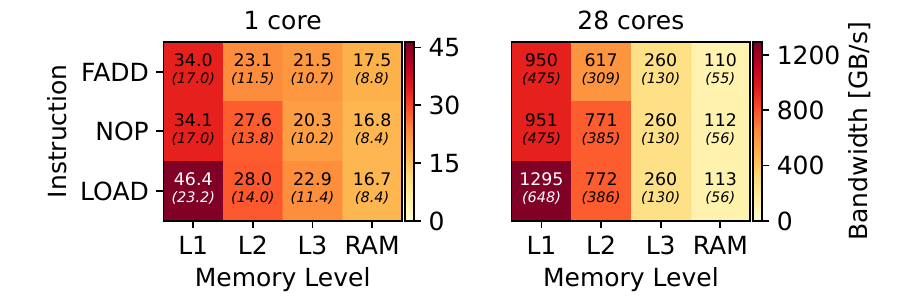}
  \caption{ThunderX2: Throughput of different NEON instructions for all memory levels using a single core and all 28 cores in [GB/s] and \textit{([B/cycle])}. The standard deviation is $<$ \SI{1}{\percent} for all cases.}
  \label{fig:thx2_single_multi}
\end{figure}

We show measurements on the ThunderX2 using the NEON extension in \figref{thx2_single_multi}.
All measurements are without SMT, as the benchmark fully saturates the execution units with a single thread.
\SI{53}{\percent} of its theoretical peak performance were reached in the \texttt{FADD} case and \SI{73}{\percent} using only load instructions.
Using \texttt{NOP} instructions instead of \texttt{FADD} did not yield any performance benefits, peaking at \SI{53}{\percent}.
A similar behavior can be observed for the A64FX in \secref{results_a64fx} and the Ampere Altra in \secref{results_aa}.
The L2 and L3 cache performance is influenced by the instruction mix.
The lowest performance was measured with \texttt{FADD} instructions, whereas the difference between only load and \texttt{NOP} is negligible.

In the single socket case with 28 cores, the L1d cache bandwidth scales as expected and in all three cases we measure $\approx 28\times$ the throughput of a single core.
The influence of the instruction mix on the L2 cache bandwidth and linear scaling with the number of cores can be observed again.
We have measured $\approx27\times$ the single core throughput using \texttt{FADD} and $\approx28\times$ with only load or \texttt{NOP} instructions.
The L3 cache does not scale linearly with the number of cores, reaching $\approx12\times$ the single core throughput for the \texttt{FADD} case.
Load or \texttt{NOP} achieve $11\times$ and $13\times$ scaling.
Our benchmark achieved \SI{66}{\percent} of the main memory peak performance, which is in line with the measurements presented by Hammond et al. \cite{thunderx2_astra_hpc}.
The measurements for both sockets are not depicted since they scaled as expected from the single socket case and did not yield any insights.

\section{Conclusion and Further Work}
\label{sec:conclusion}
Understanding the increasingly complex memory subsystems of modern processors is crucial for the analysis of the scalability and performance of applications.
The rise of Arm processors in the server and HPC space demands tools and benchmarks to evaluate their performance and microarchitectural features, similar to those available for x86-based processors.
In this work we presented such a benchmark, the Arm-membench throughput benchmark, and evaluated three processor designs.

Modern Arm processors are only capable of reaching their theoretical peak performance of the L1d cache when heavily utilizing existing SIMD extensions.
SIMD extensions with increasing register width place greater demands on the memory hierarchy.
The impact of L1d data paths, designed with the demands of SIMD in mind, was observed in the A64FX (\secref{results_a64fx}).
The A64FX has approx. $2.4\times$ the L1d cache bandwidth of the Ampere Altra (\secref{results_aa}) and $3.6\times$ the bandwidth of the ThunderX2 (\secref{results_thx2}), even though its frequency is $0.6\times$ and $0.9\times$, respectively.
This increase is solely due to the wider load and store paths from the CPU to the L1d cache, but it can only be fully utilized when using the available SIMD width.

We did observe throughput values well below the theoretical maximum of each architecture.
This is likely caused by the very nature of the load/store design of the Arm ISA.
Unlike x86, we need dedicated load instructions in addition to the arithmetic instructions whose throughput we intend to measure.
This increases the number of instructions that the processor's front end and out-of-order resources need to handle.
Therefore, an Arm processor should ideally have wider instruction fetch and decode units, as well as more out-of-order capabilities, compared to x86.
None of the processors we analyzed was able to feed enough instructions to the back end to saturate the load/store units.
This and a lack of publicly available documentation or coverage in scientific literature poses a challenge for the proper attribution of measured effects to either the microarchitecture or the benchmark itself.
Compared to other benchmarks such as Likwid and STREAM, we were able to achieve throughput results closer to the limit of the architecture (cf. \figref{a64fx_main_memory}).
We also observed more consistent results within the same memory level compared to Likwid, allowing more precise analyses of a microarchitecture.
We believe that our benchmark and analyses in this paper provide valuable input for future research in this field.

In future works we plan to analyze new and upcoming architectures with our benchmark.
We furthermore plan to port the x86-membench memory latency benchmark to Armv8 with support for MESI cache coherence states.
Memory access latency is another key metric for the cache hierarchy, affecting both local and remote cache accesses.
\pagebreak
\section*{Acknowledgments \& Reproducibility}
Research with the Fujitsu A64FX was performed on the NHR@KIT Future Technologies Partition testbed funded by the Ministry of Science, Research and the Arts Baden-Württemberg and by the Federal Ministry of Education and Research.
The source code for Arm-membench as well as a reproducibility package will be published with the final version of this paper.

\bibliographystyle{splncs04}
\bibliography{main}

\end{document}